# Lorentz transformation by mimicking the Lorentz transformation

Bernhard Rothenstein[1] and Stefan Popescu[2]

*1) Politehnica University of Timisoara, Physics Department,*
*Timisoara, Romania* brothenstein@gmail.com
*2) Siemens AG, Erlangen, Germany* stefan.popescu@siemens.com

**Abstract**. *We show that starting with the fact that special relativity theory is concerned with a distortion of the observed length of a moving rod, without mentioning if it is a "contraction" or "dilation", we can derive the Lorentz transformations for the space-time coordinates of the same event. This derivation is based on expressing the length of the moving rod as a sum of components with all the lengths involved in this summation being measured by the observers of the same inertial reference frame.*

## 1. An interpretation of the Lorentz transformation

Derived in many different ways, the Lorentz transformation (LT) for distances measured along the overlapped OX(O'X') axes of the **I** and **I'** inertial reference frames in the standard arrangement, tell us that:

$$\sqrt{1-\beta^2}\,\Delta x = \Delta x' + c\beta\Delta t' \qquad (1)$$

$$\sqrt{1-\beta^2}\,\Delta x' = \Delta x - c\beta\Delta t \qquad (2)$$

where $\beta = \dfrac{V}{c}$, V representing the relative velocity of the involved inertial reference frames.[1] In order to give (1) and (2) a physical interpretation consider the relative positions of the of **I** and **I'** as detected from I when the clocks of that frame read t ($\Delta t = t - 0$) as shown in Figure 1.

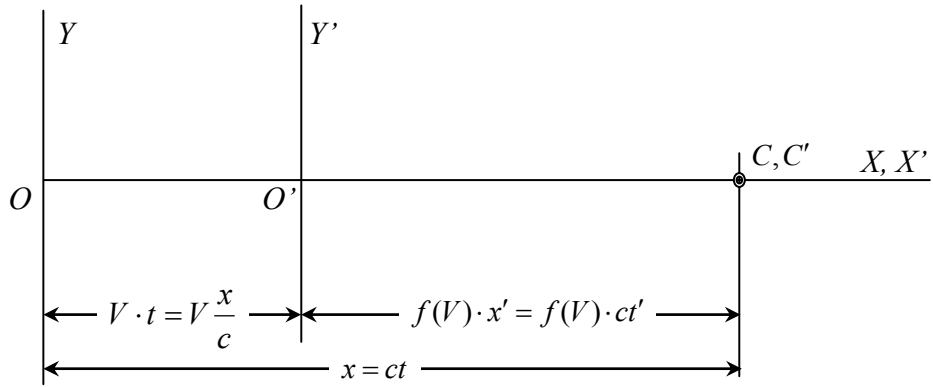

**Figure 1**. *The relative positions of the inertial reference frames **I** and **I'** in the standard arrangement as detected from I when the clocks of I read t ($\Delta t = t - 0$). All the marked distances are measured by the observers of I.*

Also consider the relative positions of the two frames when the clocks of **I'** read t'($\Delta$t'=t'- 0) as shown in Figure 2.

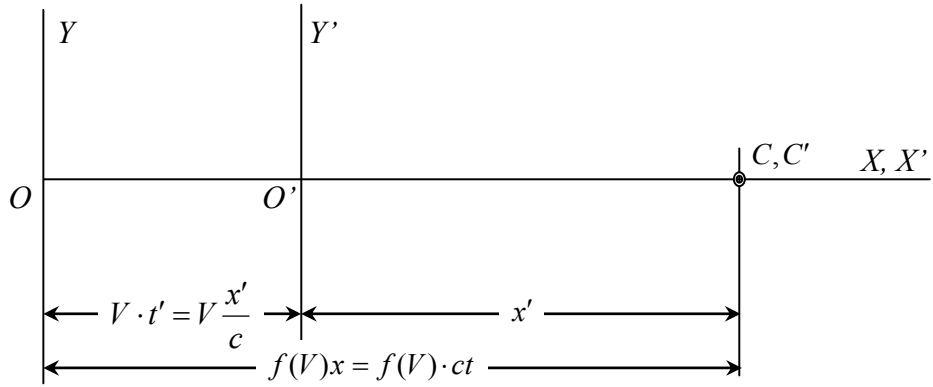

**Figure 2**. *The relative positions of the inertial reference frames **I** and **I'** in the standard arrangement as detected from **I'** when the clocks of that frame read t'=0 ($\Delta$t'=t'-0).*

The clocks of the two frames are synchronized following a clock synchronization procedure proposed by Einstein[2] ensuring that:

$$\frac{\Delta x}{\Delta t} = \frac{x-0}{t-0} = c \tag{3}$$

$$\frac{\Delta x'}{\Delta t'} = \frac{x'-0}{t'-0} = c \; . \tag{4}$$

The events involved in the transformation process are **E**(x,y=0,z=0,t) when detected from **I** and **E'**(x',y'=0,z'=0,t') when detected from **I'** where **E** and **E'** represent the **same** event because they take place at the same point in space, M(x,0,0) in **I** and M'(x',0,0) in **I'** when the clocks C(x,0,0) and C'(x',0,0) located at that point read t and t' respectively.

The relative character of length in special relativity theory makes that we can present on a **space diagram** only distances measured by the observers of the same inertial reference frame. Comparing (1) with the lengths marked in Figure 2 we see that it expresses the **Lorentz contracted length** of a rod of proper length $\Delta$x=x-0 calculated by observers of **I'** relative to which it moves with speed V as the sum of two proper lengths both measured at time t':

1) V·(t'-0)   = the distance between the origins O and O' and
2) $\Delta$x'=x'- 0 = the distance between M' and origin O'

All the lengths involved in the summation are measured by observers from **I'**. Comparing (2) with the lengths marked in Figure 1 we see that it expresses a summation of two lengths measured by observers from **I**.

The findings presented above suggest a derivation of the LT for the space-time coordinates of the same event, based on the fact that length has a relative character, without making the distinction from the beginning if length contraction or length dilation takes place.

## 2. Lorentz transformations for the space time coordinates of the same event based on the addition of lengths measured by observers of the same inertial reference frame

We start by presenting the relationship between the proper length $L_0$ of a rod (its length measured by an observer relative to whom it is in a state of rest) and its observed length L (the distorted length measured by an observer relative to whom it moves with speed V=βc) as

$$L = f(V)L_0 \tag{5}$$

where f(V) represents an unknown function of the relative speed V having the following physical properties:

- It depends only on the relative speed V and not on the proper length $L_0$, because to one rod of proper length $L_0$ at rest in one of the inertial reference frames corresponds a single length-distorted rod in another one, relative to which it moves.

- It is a relativistic invariant $f(V) = \dfrac{L}{L_0}$ i.e. it is the same in all inertial reference frames independent of the fact in which reference frame it is in a state of rest, because otherwise we could detect, confined in one of the reference frames, if we are in a state of rest or in a state of uniform motion.

Referring to Figure 1 we have

$$x = \beta ct + f(V)x' \tag{6}$$

whereas referring to Figure 2 we have

$$x' = f(V)x - \beta ct' . \tag{7}$$

Taking into account (3) and (4), equation (6) leads to

$$f(V)t' = (1-\beta)t \tag{8}$$

with equation (7) leading to

$$f(V)t = (1+\beta)t . \tag{9}$$

Multiplying (8) and (9) side by side we obtain

$$f(V) = \sqrt{1-\beta^2} \tag{10}$$

and

$$L = L_0\sqrt{1-\beta^2} \tag{11}$$

the length distortion being a contraction. With (10) equations (6) and (7) become

$$x' = \gamma(x - c\beta t) \tag{12}$$

and

$$x = \gamma(1 + c\beta t) \tag{13}$$

where $\gamma = \dfrac{1}{\sqrt{1 - \beta^2}}$. (14)

Dividing both sides of (12) and (13) with c and taking into account (3) and (4) we obtain the LT for the time coordinates

$$t = \gamma(t' + \frac{\beta}{x} x') \tag{15}$$

and

$$t' = \gamma(t - \frac{\beta}{c} x).$$

## Conclusion

Levy[3] follows a similar way starting with the formula which accounts for the Lorentz contraction (11) as known from a thought experiment which involves the light clock.[4] Our derivation avoids this knowledge being based simply on the fact that we can add or compare only lengths measured by observers of the same reference frame. It follows that the length distortion is a direct consequence of the two postulates on which special relativity is based. The explicit result is that the length distortion is in fact a contraction.